\documentstyle[aps,psfig]{revtex}
\begin{document}

\twocolumn[\hsize\textwidth\columnwidth\hsize\csname@twocolumnfalse\endcsname

\title{Electron Localization at Metal Surfaces} 

\author{L. De Santis,$^{1,2}$ and R.  Resta$^{1,3}$} \address{ $^1$INFM --
Istituto Nazionale di Fisica della Materia\\ $^2$SISSA -- Scuola
Internazionale Superiore di Stud\^\i\ Avanzati, Via Beirut 4, 34014 Trieste,
Italy \\ $^3$Dipartimento di Fisica Teorica, Universit\`a di Trieste, Strada
Costiera 11, 34014 Trieste, Italy} 

\maketitle

\begin{abstract} We investigate some surfaces of a paradigmatic $sp$ bonded
metal---namely, Al(110), Al(100), and Al(111)---by means of the ``electron
localization function" (ELF), implemented in a first--principle
pseudopotential framework. ELF is a ground--state property which
discriminates in a very sharp, quantitative, way between different kinds of
bonding. ELF shows that in the bulk of Al the electron distribution is
essentially jelliumlike, while what happens at the surface strongly depends
on packing. At the least packed surface, Al(110), ELF indicates a free--atom
nature of the electron distribution in the outer region. The most packed
surface, Al(111), is instead at the opposite end, and can be regarded as a
jellium surface weakly perturbed by the presence of the ionic cores. 


\end{abstract}

\bigskip\bigskip
]

\narrowtext

\section{Introduction}

For every given solid surface, breaking of periodicity in one dimension will
result in a change in the electronic states near and at the surface, since
the lack of nearest neighbours on one side of the surface atoms causes a
sensible local rearrangement of the surface structure and chemical bonds. We
investigate here aluminum, which as a bulk material is a paradigmatic
jelliumlike metal: cohesion and bonding are dominated by electron--gas
features, where the ions can be considered roughly speaking as a
perturbation.~\cite{pseudo} What happens at an Al surface, however, is much
less intuitive: does the surface behave essentially like a jellium surface,
or do surface atoms play a preminent role? Here we investigate this issue and
we show how a sharp answer is provided by a tool which is a very innovative
one in condensed matter physics. So far, bonding features at a crystal
surfaces have been investigated by studying either the charge density or the
local density of states. Instead, we are using here the so--called ``electron
localization function'' (ELF),\cite{Becke_90,Silvi_94,Savin_97} which has
become recently very popular in the field of quantum chemistry.

\begin{figure}[tbp]
\centerline{\psfig{figure=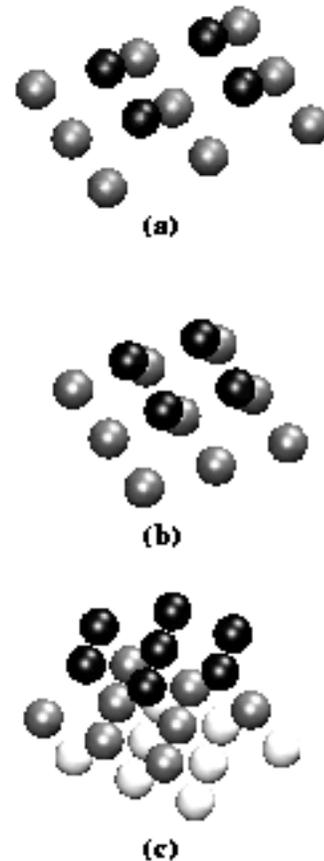}}
\caption{Schematic three--dimensional view of aluminum surfaces cut along
different crystal orientations. (a) (110) surface, (b) (100) surface, and
(c) (111) surface. Black balls correspond to topmost atoms, gray balls to
second--layer atoms, and (for the (111) surface) white balls to third--layer
atoms.}
\label{f:Fig1}
\end{figure}

We study here three basic choices for the orientation of the Al surface:
(110), (100), and (111), schematically shown in Fig.~\ref{f:Fig1}. These
high--symmetry surfaces have a rather different packing of the surface atoms:
this is also visible in Fig.~\ref{f:Fig1}, where the same scale has been used
for the three surfaces. Some correlation between packing and bonding
properties is obviously expected, but what is surprising is that the three
chosen surfaces span the whole range of possibilities, with (110) and (111)
being at the two very extreme ends: while the Al(110) surface prominently
displays well characterized Al atoms, Al(111) is essentially a weakly
perturbed jellium surface. The atomiclike vs. jelliumlike character of
the electron distribution is perspicuously shown by the immediate graphical
language of ELF.

Several appealing features make ELF the tool of choice in the present study:
ELF is a pure ground--state property, as the density is, but it ``magnifies''
by design the bonding features of a given electron distribution. Furthermore,
ELF is dimensionless, and allows to compare the nature of bonding on an
absolute scale.  ELF provides in a very simple way a quantitative estimate of
the ``metallicity'' of a given bond, or more generally of a given valence
region of the system. From our ELF calculations it is clear how in the bulk
metal the valence electrons display a free--electron nature: actually,
outside the core region, the charge density basically behaves as a
free--electron gas. This jelliumlike behaviour of the crystalline system must
be contrasted with the opposite extreme of the isolated atom, where the
valence--shell region is characterized by very different ELF values. Through
a series of ELF contour plots, we will show in the following how this
free--atom behaviour is almost entirely reproduced by the topmost atoms of
the least packed metal surface studied, namely Al(110). We stress that ELF
discriminates an atomiclike valence--electron distribution from a jelliumlike
distribution in a sharp quantitative way.

We are adopting here a fully {\it ab--initio} method, which has become the
``standard model'' in modern first--principles studies of simple metals,
covalent semiconductors, simple ionic solids, and many other disparate
materials:\cite{Cohen_91} namely, density functional theory
(DFT)\cite{Lundqvist_83} with norm--conserving
pseudopotentials.\cite{Pickett_89} This framework is the natural choice for a
real solid system as the one chosen for this work, although the ELF has been
originally introduced and mostly applied, in the quantum--chemistry
litterature, as an all--electron tool. Anyhow, we will demonstrate that the
use of the pseudopotential approximation makes the ELF particularly
meaningful, since getting rid of the core electrons and focussing solely on
the bonding electrons notably simplifies the final outcome.

In a recent very interesting paper, Fall {\it et al.}~\cite{Fall_98} have
investigated the trend in the Al work function for the same three
orientations as considered here. They provide an explanation of the trend in
terms of ``face--dependent filling of the atomiclike $p$ states at the
surface''. The ELF is an orbital--independent tool, yet it provides a
concomitant message: we show that there is indeed a face--dependent filling
of the electronic states localized in the surface region.  

The present paper is organized as follows. In Sec. II we introduce the ELF
definition, reviewing its foundamental properties and features. In Sec. III
we discuss the use of the pseudopotential approximation for our ELF
analysis. In Sec. IV we present our results carried out for the
three aluminum surfaces. Finally, in Sec. V we draw a few conclusions.

\section{Definitions}

The ELF has been originally proposed by Becke and Edgecombe,~\cite{Becke_90}
as a convenient measure of the parallel--spin electron correlation. Starting
from the short--range behavior of the parallel--spin pair probability, they
defined a new scalar function, conveniently ranging from zero to one, that
uniquely identifies regions of space where the electrons are well localized,
as occurs in bonding pairs or lone--electron pairs. This tool has immediately
shown its power to visualize the chemical bonding and the electron
localization. The ELF is defined as \begin{equation} \mbox{ELF}=\frac
1{1+[D({\bf r})/D_h({\bf r})]^2},  \label{e:ELF} \end{equation} in which
$D({\bf r})$ and $D_h({\bf r})$ represent the curvature of the parallel--spin
electron pair density for, respectively, the actual system and a homogeneous
electron gas with the same density as the actual system at point {\bf r}. By
definition, ELF is identically one either in any single--electron
wavefunction or in any two--electron singlet wavefunction: in both cases the
Pauli principle is ineffective, and the ground wavefunction is nodeless or,
loosely speaking, ``bosonic''. In a many--electron system ELF is close to one
in the regions where electrons are paired to form a covalent bond, while is
small in low--density regions; ELF is also close to one where the unpaired
lone electron of a dangling bond is localized. Furthermore, since in the
homogeneous electron gas ELF equals 0.5 at any density, values of this order
in inhomogeneous systems indicate regions where the bonding has a metallic
character.

Savin {\it et al.}~\cite{Savin_92} have proposed another illuminating
interpretation of the same quantity, showing how $D$ can be simply calculated
in terms of the local behaviour of the kinetic energy density, thus making no
explicit reference to the pair distribution function. This is particularly
convenient in our case, since the pair density is outside the scope of DFT.
Owing to the Pauli principle, the ground--state kinetic energy density of a
system of fermions is no smaller than the one of a system of bosons at the
same density: ELF can be equivalently expressed in terms of the extra
contribution to the kinetic energy density due to the Pauli principle.  The
Savin {\it et al.} reformulation also provides a meaningful physical
interpretation: where ELF is close to its upper bound, electrons are strongly
paired and the electron distribution has a local ``bosonic'' character.

Following Ref.~\onlinecite{Savin_92}, $D({\bf r})$ is the Pauli excess
kinetic energy density, defined as the difference between the kinetic energy
density and the so--called von Weizs\"{a}cker kinetic energy
functional:~\cite{Dreizler} \begin{equation} D({\bf r})=\frac 12\nabla _{{\bf
r}}\nabla _{{\bf r^{\prime }}}\rho ({\bf r}, {\bf r^{\prime }})\bigg |_{{\bf
r}={\bf r^{\prime }}}-\frac 18\frac{|\nabla n({\bf r})|^2}{n({\bf r})}
\label{e:D} \end{equation} where $\rho $ is the one--body reduced
(spin--integrated) density matrix.  The von Weizs\"{a}cker functional
provides a rigorous lower bound for the exact kinetic energy
density~\cite{Dreizler} and is ordinarily indicated as the ``bosonic''
kinetic energy, since it coincides with the ground--state kinetic energy
density of a non--interacting system of bosons at density $n({\bf r})$. 
Therefore, $D$ is positive semidefinite and provides a direct measure of the
local effect of the Pauli principle. The other ingredient of
Eq.~(\ref{e:ELF}) is $D_{h}({\bf r})$, defined as the kinetic energy density
of the homogeneous electron gas at a density equal to the local density:
\begin{equation} D_h({\bf r})=\frac 3{10}(3\pi ^2)^{\frac 23}n({\bf
r})^{\frac 53}.  \label{e:D_h} \end{equation} It is obvious to verify that
the ELF is identical to one in the ground state of any one-- or two--electron
systems, while is identical to 0.5 in the homogeneous electron gas at any
density.

As commonly done in many circumstances---including in other ELF
investigations\cite{Savin_92}---we approximate the kinetic energy of the
interacting electron system with the one of the noninteracting Kohn--Sham
(KS) one. We therefore use the KS density matrix: \begin{equation} \rho ({\bf
r},{\bf r^{\prime }})= 2 \sum_i\phi_i^{KS} ({\bf r})\phi _i^{*KS}( {\bf
r^{\prime }}),  \label{e:rho} \end{equation} where $\phi _i^{KS}({\bf r})$
are the occupied KS orbitals. Such approximation is expected to become
significantly inaccurate only in the case of highly correlated materials.

\section{Isolated Pseudo--Atom}

The original ELF definition is an all--electron one, and has the remarkable
feature of naturally revealing the entire shell structure for heavy atoms. 
Such a feature is of no interest here, since only one electronic shell (the
$sp$ valence one) is involved in the bonding of Al atoms in any
circumstances. The pseudopotential scheme simplifies the landscape, since
only the electrons of the relevant valence shell are dealt with explicitly:
the ELF message comes out therefore much clearer, with basically no loss of
information.\cite{Kohout_97} In the spatial regions occupied by core
electrons, the pseudo--electronic distribution shows a depletion and ELF
assumes very low values. Outside the ionic cores, in the regions relevant to
chemical bonding, the norm conservation endows the pseudocharge density with
physical meaning, as widely discussed in the modern pseudopotential
literature.\cite{Cohen_91,Pickett_89} The pseudo ELF carries therefore---in
the material of interest here---the same information as the all--electron
ELF, while it removes irrelevant and confusing features due to the inner,
chemically inert, shells.

\begin{figure}[tbp]
\centerline{\psfig{figure=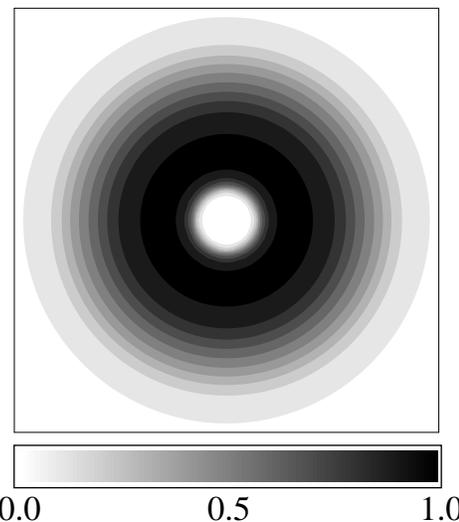}}
\caption{ELF for an isolated {\it pseudo}--atom. The grey--scale is also
shown: dark (clear) regions correspond to large (small) ELF values. The same
grey--scale is adopted in all subsequent figures.}
\label{f:Fig2}
\end{figure}

For an isolated pseudo--atom we have by construction only a single valence
shell, clearly displayed by a single and very prominent ELF maximum. In
Fig.~\ref{f:Fig2} we report our result carried out for an isolated Al atom:
in the picture one immediately remarks the spherical region associated to the
ELF maximum (0.91), where the charge density is almost bosonic. The black
cloud in Fig.~\ref{f:Fig2} indicates the strong localization of the
electronic valence shell and perspicuously distinguishes the {\it free}
aluminum pseudo--atom from the crystalline one. 

Indeed looking at the following Figs.~\ref{f:Fig3}--\ref {f:Fig5}, and
inspecting the bulk regions of them, the ELF plots quantitatively
demonstrate that the valence electrons have a predominant free--electron
nature, with a jelliumlike (or Thomas--Fermi) ELF value. The ion cores only
provide an ``exclusion region'' for electrons (white circles in the plots),
but basically do not alter the jelliumlike nature of the electron
distribution outside the core radii: in this sense we may regard the ion
cores as a ``weak'' perturbation.\cite{pseudo} Actually, the maximum value
attained by ELF between nearest--neighbor atoms in bulk Al is only 0.61
and---outside the core radii---there is a widespread grey region, where ELF
is almost constant and close to 0.5, thus indicating a jelliumlike
electronic system.  This peculiar behaviour will help us in the following
to understand the bonding pattern occurring at different Al surfaces.

\section{Aluminum Surfaces}

All the calculations in this work use a state--of--the--art set of
ingredients: a plane--wave expansion of the KS orbitals with a 16 Ry
kinetic--energy cut--off, a set of Monkhorst--Pack\cite{Monkhorst_76} special
points for the Brillouin zone integration, with a Gaussian broadening of
0.01Ry and a norm--conserving pseudopotential in fully non--local
form.\cite{Kleinman_82} The calculations for the (111) surface were performed
using a 9+6 supercell (9 planes of Al, 6 equivalent planes of vacuum) and 37
{\bf k}--points in the irreducible Brillouin zone. The corresponding values
used for the (100) surface are an 8+6 supercell and 46 {\bf k}--points, and
for the (110) surface an 8+8 supercell and 48 {\bf k}--points. We have
preliminarily investigated the effect of surface ionic relaxation on ELF, and
found it negligible. Similar insensitiveness was found in
Ref.~\onlinecite{Fall_98} upon other surface electronic properties: we
therefore present results for the unrelaxed surfaces. As shown above in
Fig.~\ref{f:Fig1}, the three surfaces have a different packing: the
packing increases from Al(110) to Al(100) to Al(111). A measure of this
packing is the number of nearest neighbors in the surface plane, called
coordination in the following: this number is 2, 4, and 6 for the three
surfaces, respectively.

\begin{figure}[tbp]
\centerline{\psfig{figure=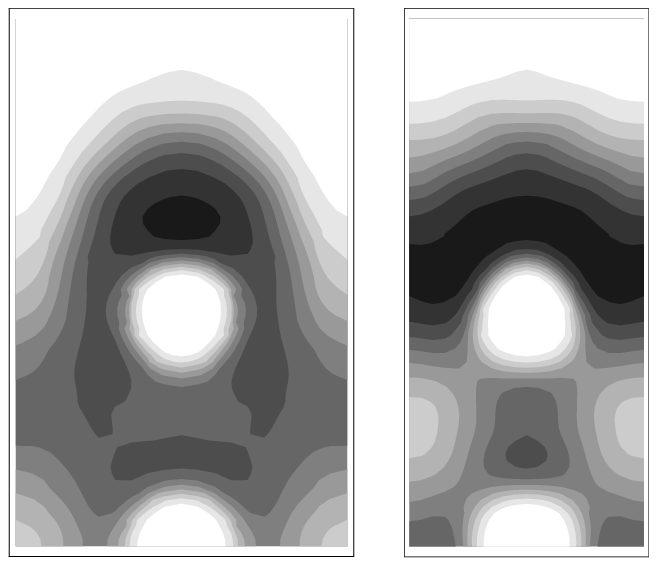}}
\caption{Al(110): ELF contour plot in two non--equivalent planes containing
the topmost atom and orthogonal to the surface.}
\label{f:Fig3}
\end{figure}

\begin{figure}[tbp]
\centerline{\psfig{figure=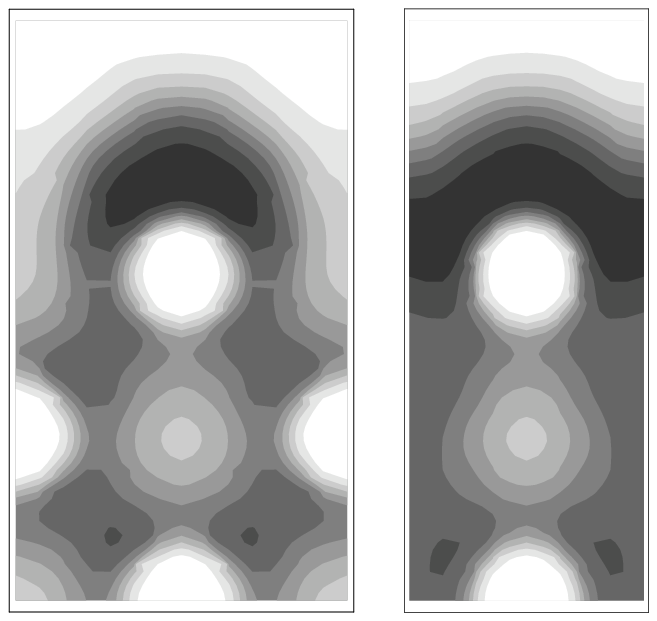}}
\caption{Al(100): ELF contour plot in two non--equivalent planes containing
the topmost atom and orthogonal to the surface.}
\label{f:Fig4}
\end{figure}

In Fig.~\ref{f:Fig3} we report two ELF contour plots along two
non--equivalent planes passing through the topmost (twofold coordinated)
atom of the Al(110) surface. Along a direction orthogonal to the surface and
passing through the surface atom, our calculated ELF attains the maximum
value of 0.86, thus indicating that the Pauli principle has little effect. 
Comparing Fig.~\ref{f:Fig3} to \ref{f:Fig2}, where the ELF maximum has the
close value of 0.91, we quantitatively see see how much the surface atom
actually behaves as a free--atom in the outer direction. A slightly different
electronic distribution is present at the four--fold coordinated Al(100)
surface in Fig.~\ref{f:Fig4}. In this case the ELF maximum---along the same
direction---attains the lower value of 0.80 and shows therefore an electron
distribution with a less pronounced atomiclike character.

The almost black regions around the surface atoms in Figs.~\ref{f:Fig3} and
\ref{f:Fig4} are also clearly visible in the top views of Fig.~\ref{f:Fig6},
where the contour plots are drawn in the outermost atomic plane. In this
plane the maxima occur midway between nearest-neighbors atoms and assume the
values of 0.77, 0.74, and 0.67 in order of increasing packing. This decrease
of electron pairing shows a trend from covalent--like to metallic--like
bonding between nearest neighboring surface atoms; bonding to the underlying
bulk atoms is instead metallic in all cases.

Finally, in Fig.~\ref{f:Fig5} we have the extreme case of the six--fold
coordinated Al(111) surface. The absolute ELF maximum is only 0.73 and lies
in a low--symmetry point, slightly off the outermost atomic plane. Along the
direction orthogonal to the surface and passing through the surface atom, the
ELF maximum is only 0.65: this behaviour clearly marks an almost smooth decay
from the bulk---where the electrons exhibit a prevailing jelliumlike
distribution---to the vacuum region. Also in the top view of
Fig.~\ref{f:Fig6} there is no evidence of electron pairing between surface
atoms, with a wide spread region of almost uniform ELF, close in value to
0.5. We can therefore summarize the result by saying that the Al(111) surface
is essentially a jellium surface perturbed by the atomic cores, in a
pseudopotential sense:\cite{pseudo} the perturbation obviously originates an
``exclusion region'' (white plots within the core radii), but besides this
has little effect on the electron distribution in the bulk {\it and} at the
(111) surface.

The values of the maxima in the direction orthogonal to the surface and
passing through the surface atom in the three cases (0.91, 0.80, and 0.65, as
reported above) can be interpreted as a measure of the occupation of the
atomiclike states protruding from the surface. We recall in fact that ELF is
identically equal to 1 for any one-- or two--electron system: high ELF values
indicate a strong localization of the wavefunction. Our numbers show an
analogous trend as the one pointed out by Ref.~\onlinecite{Fall_98}, where it
is shown that the filling of atomic $p_\perp$ orbitals, protruding from the
surface, decreases indeed with increasing packing, {\it i.e.} going from
(110) to (100) to (111).

\section{Conclusions}

In conclusion, we have shown how the ELF at the surfaces of a paradigmatic
$sp$--bonded metal distinctly reveals the rearrangements in the electron
distribution due to the surface formation and to the changes in the local
coordination. We have shown that ELF, in a modern DFT--pseudopotential
framework, is a powerful tool to investigate---without using any spectral
information---the chemical bonding pattern: this makes the method appealing
in order to deal with much more complicated systems.  

Finally, to put this work in a wider perspective, we notice in the current
literature a quest for other innovative tools which address localization
and bonding,\cite{Silvestrelli98,rap107} and whose relationships to ELF
have not been investigated yet.

\begin{figure}[tbp]
\centerline{\psfig{figure=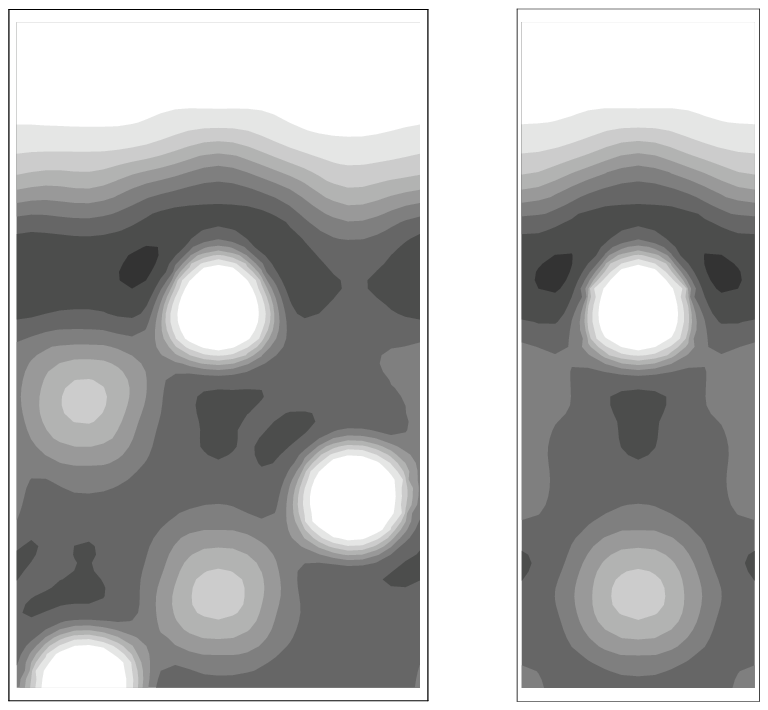}}
\caption{Al(111): ELF contour plot in two non--equivalent planes containing
the topmost atom and orthogonal to the surface.}
\label{f:Fig5}
\end{figure}

\begin{figure}[tbp]
\centerline{\psfig{figure=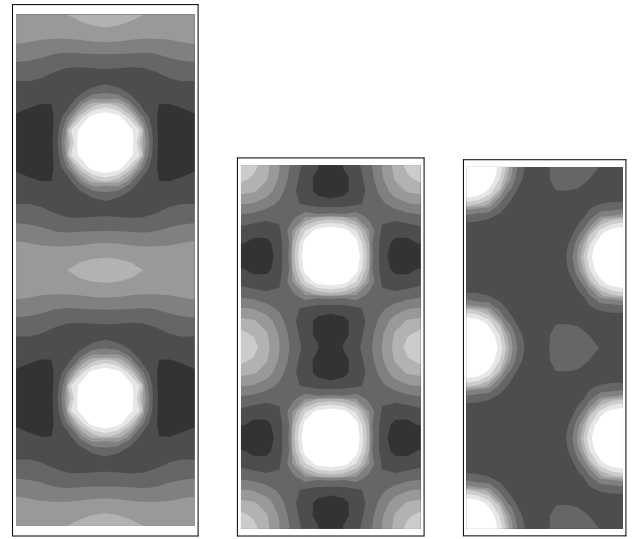}}
\caption{ELF contour plot in the plane containing the topmost Al atoms for
the three different orientations.}
\label{f:Fig6}
\end{figure}

\end{document}